\documentclass[onecollarge,natbib]{svjour2}
\bibpunct{[}{]}{;}{n}{}{,} 
\smartqed  
\usepackage{graphicx}
\usepackage{amssymb}
\usepackage{amsmath}
\usepackage{hyperref}
\usepackage{bm}
\usepackage{etoolbox}
\usepackage{lmodern}
\usepackage{slashed}
\usepackage{float}
\usepackage{subfig}
\usepackage{amssymb}
\usepackage{latexsym}
\usepackage{feynmp}
\usepackage{wrapfig}
\journalname{Few-Body Systems}
\newcommand{\mm}{\mu\bar{\mu}}
\newcommand{\ee}{e\bar{e}}
\newcommand{\tta}{\tau\bar{\tau}}
\DeclareMathOperator{\de}{\mathrm{d}}

\DeclareMathOperator{\llg}{\ell^+\ell^-\gamma}

\DeclareMathOperator{\gag}{\gamma\gamma}
\DeclareMathOperator{\gp}{\gamma^+}

\begin{document}

\title{Nonperturbative True Muonium on the Light Front with TMSWIFT \thanks{Work supported by the National Science Foundation under 
Grant Nos. PHY-1068286 and PHY-1403891 and by the International Light Cone Advisory Committee under the McCartor Fellowship program.}
}


\author{Henry Lamm         \and
        Richard F. Lebed 
}


\institute{Henry Lamm \at
               Department of Physics, Arizona State University, P.O. Box 871504, Tempe, AZ 85287-1504 USA.\\
              \email{hlammiv@asu.edu}           
           \and
           Richard F. Lebed \at
	    Department of Physics, Arizona State University, P.O. Box 871504, Tempe, AZ 85287-1504 USA.\\
	  \email{Richard.Lebed@asu.edu}
}

\date{Received: date / Accepted: date}

\maketitle

\begin{abstract}
The true muonium $(\mm)$ bound state presents an interesting test of 
light-cone quantization techniques. In addition to exhibiting the standard problems of 
handling non-perturbative calculations, true muonium requires correct 
treatment of $\ee$ Fock-state contributions. Having previously produced a 
crude model of true muonium using the method of iterated resolvents, our current 
work has focused on the inclusion of the box diagrams to improve the 
cutoff-dependent issues of the model. Further, a parallel computer code, TMSWIFT, allowing 
for smaller numerical uncertainties, has been developed. This work focuses on 
the current state of these efforts to develop a model of true muonium that is testable 
at near-term experiments.
\end{abstract}

\section{Introduction}
The current state of flavor physics has a ``muon 
problem''.  Several muon observables 
\cite{PhysRevD.73.072003,Antognini:1900ns,Aaij:2014ora,CMS:2014hha} show disagreement with Standard Model calculations.  A strong candidate for providing clues to resolve this problem is the bound 
state $(\mm)$, dubbed ``true muonium''\cite{Hughes:1971}.  Traditional bound states have limited new physics reach due to small reduced masses $\mu\approx m_e$ or nuclear-structure uncertainties. In contrast, true muonium's $\mu=m_\mu/2$ and leptonic nature make it an ideal probe, through the Lamb shift or hyperfine splitting\cite{TuckerSmith:2010ra,PhysRevD.91.073008,Lamm:2015fia,Lamm:2015gka}.

True muonium has not been observed due to 
difficulties in producing associated low-energy muon pairs and 
its short lifetime ($\tau\approx$ 1 ps).  Many proposed methods of 
production channels exist\cite{Nemenov:1972ph,Moffat:1975uw,Holvik:1986ty,Kozlov:1987ey,Ginzburg:1998df,
ArteagaRomero:2000yh,Brodsky:2009gx,Chen:2012ci,Banburski:2012tk,Ellis:2015eea}.
The Heavy Photon Search (HPS) experiment plans to search for true muonium in 2016\cite{Celentano:2014wya,Banburski:2012tk}.  Additionally, the DImeson 
Relativistic Atom Complex (DIRAC) could observe it in an upgraded run\cite{Benelli:2012bw,dirac}. Given enough statistics, DIRAC could measure the Lamb shift using methods developed for ($\pi\bar{\pi}$) \cite{Nemenov:2001vp}.  These experiments produce relativistic true muonium.   Unfortunately, instant-form (conventional fixed time) wave functions are not boost-invariant; thus production and decay rates are modified.  To reduce this uncertainty, we are producing boost-invariant wave functions through light-front techniques\cite{Lamm:2013oga}.

Quantizing at light-front time $x^+ \equiv t + z$ (called {\it front
  form})~\cite{Dirac:1949cp} allows one to develop a Hamiltonian formalism~\cite{Brodsky:1997de}. In this formalism, a analogue of the Schr\"{o}dinger
equation exists as an infinite but denumerable set of
coupled integral equations.  One
may then truncate the equations by limiting the set of Fock
states included and discretizing momenta.  This finite problem can then be solved by a computer.  This technique is called Discretized Light-Cone
Quantization (DLCQ)~\cite{Pauli:1985ps}.

Our work is a direct continuation of the
DLCQ methods developed in Ref.~\cite{Tang:1991rc,Krautgartner:1991xz,Kaluza:1991kx,Trittmann:1997xz} to true muonium.
The explicit $|\gamma\rangle$ component allows mixing between $|\ee \rangle$, $|\mm \rangle$, and $|\tta\rangle$ Fock
states.  We also implemented a counterterm subtraction to
the interaction amplitudes that improves the ultraviolet
behavior.
Following Ref.~\cite{Trittmann:1997xz}, we adopted
the larger value $\alpha = 0.3$.  In this regime, QED perturbative calculations potentially become unreliable.  We use this strong coupling value of $\alpha$ to study flavor-mixing and regularization dependence. In this work, we also investigate the approach to the physical coupling constant of QED.

\section{True Muonium model}
In front-form, the eigenvalue equation for a bound state is given by:
\begin{equation}
\label{eq:ham}
\left(M^2 - \sum_i\frac{m^2_i+\bm{k}^{2}_{\perp i}}{x_i} \right)
\psi (x_i,\bm{k}_{\perp i};h_i) =\sum_{h_j}\int_D\mathrm{d}x'_{j} \mathrm{d}^2 \bm{k}'_{\perp j}
\langle x_i,\bm{k}_{\perp i};h_i \left| V
\right| x'_{j}, \bm{k}'_{\perp j};h_j \rangle \psi (x'_{j},\bm{k}'_{\perp,j}; h_j),
\end{equation}
where $M$ is the invariant mass of the state, $m$ indicates a mass term, $i,j$ are particle indices, $x$ and $\bm{k}_\perp$ are the conventional light-front coordinates, $h$ is shorthand for all intrinsic quantum numbers of a state, and $V$ are interaction terms given by the light-front Hamiltonian.  The domain $D$ in
Eq.~(\ref{eq:ham}) is defined by introducing a cutoff $\Lambda$, and we choose~\cite{Lepage:1980fj}
\begin{equation}
\frac{ m^2 + \bm{k}_\perp^2 }{x(1-x)} \le \Lambda^2 + 4m^2 \, .
\end{equation}  

In our model, we are considering the charge-zero, lepton family number-zero states in the truncated Fock space of 
 \begin{equation}
 \label{eq:wave}
  |\Psi \rangle  =  \psi_{\mm} |\mm\rangle + \psi_{\ee} |\ee\rangle + \psi_{\tta} |\tta\rangle+
  \psi_\gamma|\gamma \rangle + \psi_{\mm\gamma}
  |{\mm\gamma} \rangle + \psi_{\ee\gamma}|\ee\gamma\rangle + \psi_{\tta\gamma} |\tta\gamma\rangle\,
\end{equation}

Solving for the eigenstates of $H_{\rm LC}$ with this limited Fock
space nonetheless gives the bound states of positronium $(\ee)$,
true muonium $(\mm)$, and true tauonium $(\tta)$, as well as associated continuum states (up to effects from neglected
higher-order Fock states).   The wave functions are in the form of
Eq.~(\ref{eq:wave}) with helicity states for only $\left|\mm\right>$,
$\left|\ee\right>$, and $\left|\tta\right>$ components. The $\left|\gamma\right>$ and $\left|\ell\bar{\ell}\gamma\right>$
components are folded into $V_{\rm eff}$ by means of the method of
iterated resolvents\cite{Pauli:1997ns,Trittmann:1997xz}.

It has been shown\cite{Krautgartner:1991xz,Trittmann:1997xz,Lamm:2013oga} that strong dependence in ${}^1S_0$
states on $\Lambda$ arises from the matrix element
between antiparallel-helicity states (called $G_2$ in App.~F.3
of Ref.~\cite{Trittmann:1997xz}).  This element approaches a constant as
$k_\perp \equiv |\bm{k}_\perp|$ or $k'_\perp \equiv |\bm{k}'_\perp|\to
\infty$.  The result is a $\delta$ function-like behavior in
configuration space.  Ref.~\cite{Krautgartner:1991xz} chose to regularize this singularity by deleting
the entire term.  We instead subtract
only its limit as $k_\perp$ or $k'_\perp \to \infty$, retaining
part of the term (including $x$ and $x'$ dependence).  This scheme removes the strong $\Lambda$
dependence of ${}^1S_0$ states in both QED\cite{Lamm:2013oga,Wiecki:2014ola} and QCD\cite{Li:2015zda} models.  At the cost of unknown regularization dependence, the model allows for taking the $\Lambda\rightarrow\infty$ limit.  We found that the $1^1S_0$ state eigenvalue can be fit by
\begin{equation}
\label{eq:fit}
 M^2(N,\Lambda)=M^2_{\infty}(1-be^{-cN})(1-de^{-f\Lambda}).
\end{equation}
\section{Results from Mesonix \& TMSWIFT}
Our initial results\cite{Lamm:2013oga} were for the system of electrons and muons.  Using a modified version of the Mesonix code\cite{Trittmann:1997xz}, we produced the entire spectrum of positronium and true muonium.  The largest $N$ we explored with this serial code was $N_\mu=N_e=37$. This required a full day of CPU time on a 3.0 GHz core to produce a single $J_z$ spectrum.  These results agreed with the instant-form corrections to the annihilation channel from electron loops, albeit with a large uncertainty due to difficulty in properly sampling continuum electron states. 

\begin{figure}
  \begin{center}
    \includegraphics[width=\linewidth]{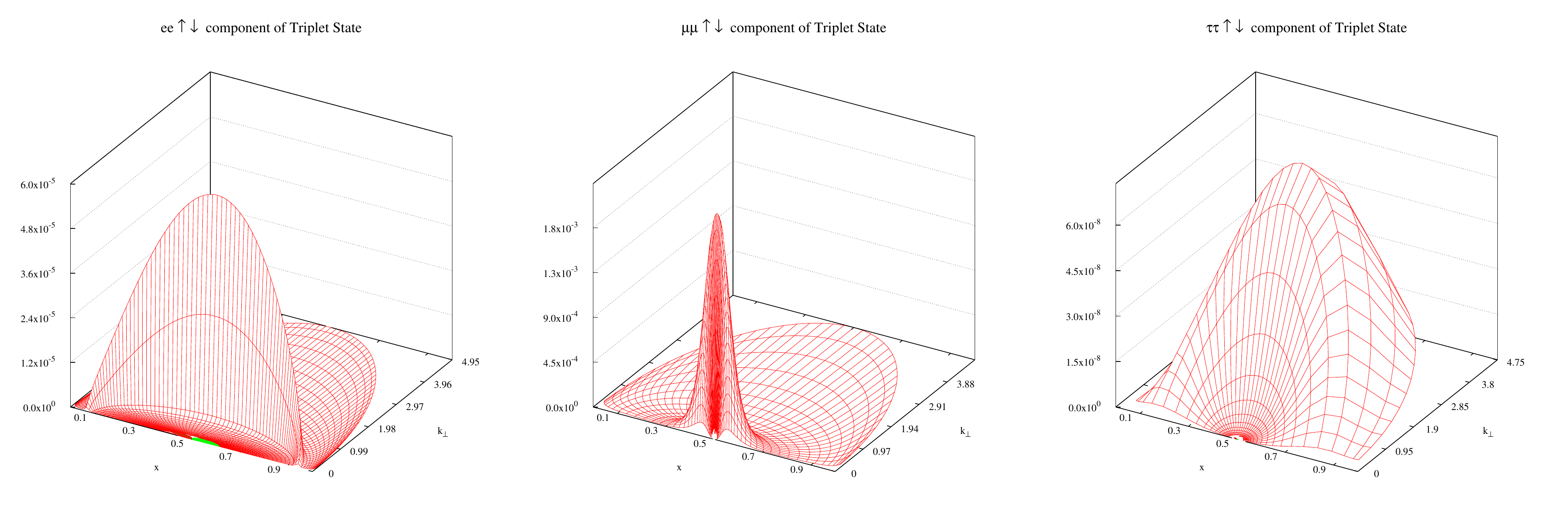}
      \caption{\label{fig:waves}The $1^3 \! S_1^0$ probability density of (left) $\uparrow\downarrow$ $\ee$, (center) $\uparrow\downarrow$ $\mm$, and (right) $\uparrow\downarrow$ $\tta$ components of true muonium with $J_z=0$, as functions of $x$ and
  $k_\perp$, for $\alpha=0.3$, $m_e=\frac{1}{2}m_\mu$, $m_\tau=2m_\mu$
  $\Lambda_i=10\alpha m_i/2$, and $N_\mu=N_\tau=37$,$N_e=71$.}
  \end{center}
\end{figure}

\begin{wraptable}[20]{r}{6cm}
\begin{center}
\begin{tabular}{l|c|c}
\hline\hline
$\alpha$&$M^2(1^1S_0)_{LF}$&$M^2(1^1S_0)_{IF}$\\ \hline
0.3&3.8944(3)&3.9053\\
0.1&3.98988(2)&3.98994\\
0.07&3.99507088(8)&3.99508575\\
0.05&3.99749246(8)&3.99749629\\
0.02&3.9996087(2)&3.99959990\\
0.01&3.999900017(5)&3.999899994\\
\hline
\end{tabular}
\caption{$M^2$ of the true muonium ground state $1^1S_0$ for a range of $\alpha$ in units of $m_\mu$.  The first column is from TMSWIFT and the second column presents $\mathcal{O}(\alpha^4)$ instant-form calculations.  The light-front results were obtained by fitting a set of calculations with $N_\mu=(13,71)$ and $\Lambda_\mu=(1,30)$ in units of $\alpha m_\mu/2$ to Eq.~(\ref{eq:fit}).  The reported errors are only from fitting.}
\label{tab:el}
\end{center}
\end{wraptable}
We have written a new parallel code, TMSWIFT (True Muonium Solver With Front-form Techniques), to overcome the numerical limitations.  This code is more flexible, as well as using the parallel eigenvalue solver package SLEPc\cite{Hernandez:2005:SSF}.  TMSWIFT allows an arbitrary number of flavors each specified by a mass, $\Lambda$, and $N$.  Different discretization schemes are available for exploration of numerical errors and efficiency.  TMSWIFT allows easy implementation of new effective interactions (e.g., from $|\gamma\gamma\rangle$ states).  We have performed three-flavor calculations with $N_{\mu,\tau}=37$ and $N_e=71$ for $J_z=0$ in 2 hours using 512 cores of the Stampede supercomputer.  In Fig.~\ref{fig:waves}, we present anti-parallel helicity wave-function components in the $1^3S_0$ state of true muonium.

Calculations with lower $\alpha$ were also performed. The numerical stability becomes worse at decreasing $\alpha$ for fixed $N$. The lowest achieved value is $\alpha=0.01\approx1.4\alpha_{\rm QED}$.  Fitting multiple values of $N$ and $\Lambda$, we were able to extract values of $M^2_\infty(1^1S_0)$ and compare them to instant-form calculations at $\mathcal{O}(\alpha^4)$ (Tab.~\ref{tab:el}).  Currently the source of the small discrepancy is undetermined, but two sources to investigate are regularization dependence and the mismatch of higher-order corrections between the $\alpha-$perturbative instant-form and the nonperturbative Fock-state front-form calculations.  
\section{The $|\gamma\gamma\rangle$ Fock State}
\begin{wrapfigure}[13]{l}{7.0cm}
\begin{center}
\includegraphics[width=0.9\linewidth]{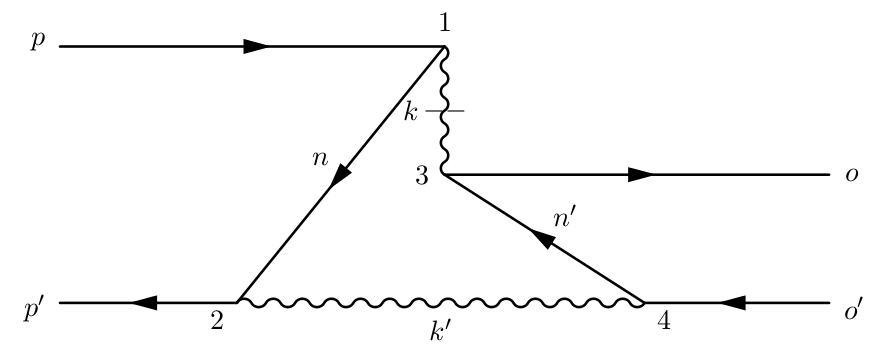}
\end{center}
\caption{\label{fig:1243} Example of a $|\gag\rangle$ intermediate-state diagram that must be computed.}
\end{wrapfigure}

We are currently working to include the $|\gamma\gamma\rangle$ state, which dominates the decay of singlet
states of true muonium (and in particular should have a pronounced
effect on ${}^1 \! S_0$ wave functions).  To do this, we have computed the sum of 19 time-ordered light-front diagrams (Fig.~\ref{fig:1243} and permutations of the vertices, plus diagrams where each internal particle is instantaneous), which include some
$|\ell\bar{\ell}\ell\bar{\ell}\rangle$ and $|\ell\bar{\ell}\ell'\bar{\ell}'\rangle$ diagrams to preserve gauge invariance.  The diagrams result in two integrals for the effective interactions, for the $|\gamma\gamma\rangle$ sector, we have
\begin{align}
 H_{\gag}=-\frac{\alpha^2}{(2\pi)^4}\int\de^3k&\left(\frac{G_{\llg}G_{\gag}G_{\llg}'}{\sqrt{p^+o^+(o')^+(p')^+}}\right)\frac{\bar{u}_1\gamma^\mu \bigg(\slashed{l}+m+\frac{\gp}{2 G_{\llg}}\bigg)\gamma^\sigma v_2\bar{v}_4\gamma^\rho\bigg(\slashed{l}'+m+\frac{\gp}{2 G_{\llg}'}\bigg)\gamma^\nu u_3}{|k^+||P^+-k^+||p^+-k^+||o^+-k^+|}\nonumber\\&\times\bigg(d_{\mu\nu}d_{\sigma\rho}\Theta_k\Theta_{-k}+\frac{\eta_{\mu\nu}d_{\sigma\rho}}{G_{\gag}|k^+|}\Theta_k\Theta_{-p}\Theta_{-o}-\frac{d_{\mu\nu}\eta_{\sigma\rho}}{G_{\gag}|(P-k)^+|}\Theta_{-k}\Theta_{p}\Theta_{o}\bigg),
\end{align}
and a similar integral occurs for the $|\ell\bar{\ell}\ell\bar{\ell}\rangle$ sector.  In this integral, $l^\mu$ and $(l')^\mu$ are not free momenta but fixed functions of the external momenta and $k^\mu$.  We further used the notation
\begin{equation}
 \Theta_k=\theta(k^+),\phantom{XXX}
 \Theta_{-k}=\theta((P-k)^+),\phantom{XXX}
 \Theta_{\pm i}=\theta(\pm (i-k)^+),
 \end{equation}
where $P^+$ is the total $+$ component at any time, $i$ indicates an external momentum, and $k$ is the $k^{\mu}$ photon's $+$ component.  We introduce $\eta_{\mu\nu}$, which is zero except for the $\eta_{++}=1$ component, and $d^{\mu\nu}(k)$ is the standard light-front photon propagator (Eq.(A.23) in Ref.~\cite{Brodsky:1997de}).  

In these equations, the non-perturbative propagators $G_i$ need to be defined such that the instantaneous diagrams are canceled correctly in the vein of Ref.~\cite{Pauli:1997ns}.  While in the $|\ell\bar{\ell}\gamma\rangle$ sector only one $G_i$ occurs, the mixing between $|\gamma\gamma\rangle$ and $|\ell\bar{\ell}\ell\bar{\ell}\rangle$ has presented unique problems in ensuring the proper cancellations.  At present we are looking at different forms of this function.  Additionally, since we include a pure photon state, we must account for the IR features of the interaction.  The simplest method to treat this would be to include a photon mass to the interaction and then study the limit as $m_\gamma\rightarrow 0$.  From these results, we can compute helicity matrix elements between Fock states and then implement them in TMSWIFT.
\section{Summary and Conclusion}
In this paper, we have computed the light-front wave functions of true muonium. This was done in a three-flavor model of QED with only 7 explicit Fock states. Our new, parallelized code is capable of simulating such larger Fock spaces.  We have shown that extrapolations to $\Lambda\rightarrow \infty$ and $N\rightarrow\infty$ are possible.  Finally, TMSWIFT also allows for the study of lower values of $\alpha$.  We are currently working to include the $|\gamma\gamma\rangle$ state and the pair of states
$|\ell\bar{\ell}\ell\bar{\ell}\rangle$ and $|\ell\bar{\ell}\ell'\bar{\ell}'\rangle$, which are required for gauge invariance.  These corrections are crucial for precision true muonium predictions and are a necessary step for QCD bound states as well.  To integrate
these states into a model, proper renormalization of the Hamiltonian
may be required.  Limitations on the size of the Fock space have been greatly decreased via TMSWIFT. This allows for more
explicit Fock-state renormalization methods like Pauli-Villars
regulators~\cite{Chabysheva:2009vm,Chabysheva:2010vk,Malyshev:2013eca} and sector-dependent
counterterms~\cite{Karmanov:2008br,Karmanov:2012aj}.  The increased speed also makes Hamiltonian-flow approaches~\cite{Gubankova:1998wj,Gubankova:1999cx} more amenable.  Using the exchange properties of leptons could reduce the number of basis states, similar to Ref.~\cite{Chabysheva:2014rra}.
\begin{acknowledgements}
This work was supported by the National Science Foundation under Grant Nos. 
PHY-1068286 and PHY-1403891 and through the McCartor Fellowship program from the International Light-Cone Advisory Committee. This work used the Extreme Science and Engineering Discovery Environment (XSEDE), which is supported by National Science Foundation grant number ACI-1053575. The authors would like to express our gratitude
to the organisers of the Light Cone 2015.
\end{acknowledgements}

\bibliographystyle{apsrev4-1}
\bibliography{wise}   
\end{document}